# Plasma Radiation Model of Fast Radio Bursts from Magnetars

Edison Liang, Rice University, Houston, TX, USA 77005

## Abstract

We propose a novel idea for the coherent intense millisecond radio emission of cosmic fast radio bursts (FRBs), which have recently been identified with flares from a magnetar. Motivated by the conventional paradigm of Type III solar radio bursts, we will explore the emission of coherent plasma line radiation at the electron plasma frequency and its harmonic as potential candidates of the coherent FRB emissions associated with magnetar flares. We discuss the emissions region parameters in relativistic strongly magnetized plasmas consisting of electrons, positrons and protons. The goal is to make observable predictions of this model to confront the multi-wavelength observations of FRBs from magnetars. These results will impact both observational radio astronomy and space-based astrophysics.

## 1. Background and motivation

FRBs are short (~ms) coherent intense radio burst (hundreds of MHz - few GHz) emitted by sources mostly at cosmological distances, whose origin(s) intrigued astronomers since its discovery in 2007 [1] [2] [3]. Over 700 FRBs have been reported as of 2023. The recent detection of a pair of FRBs from the Galactic magnetar SGR1935+2154 (CHIME/FRB collaboration, [4]) has firmly established that *at least a subset of FRBs are associated with magnetar flares*. While these two magnetar bursts are slightly sub-luminous compared to most extragalactic FRBs, and therefore may have escaped detection at typical extragalactic distances ($\geq$ 10 Mpc. Fig.1, [4]), they confirm that magnetars are definitely capable of emitting FRBs with energy output up to the most energetic FRBs. For example the energies of the two SGR1935+2154 FRBs are within a factor of a few hundred of the weakest burst of repeater FRB 121102, which is likely a magnetar. The dynamic range of FRB 121102 is $\sim 10^4$ (Fig.1). Hence if SGR1935+2154 bursts have similar dynamic range, its most energetic bursts will certainly fall in the middle of extragalactic FRBs. Due to the limited bandwidth of most radio telescopes in the GHz band, the true spectral range of FRBs is impossible to determine at present. However, at least a subset of FRBs, such as those from the repeating FRB 121102, are *observed to be narrow-band* (Fig.2), *with $\Delta w/w_o < 0.1$*, $\Delta w$=FWHM of spectral peak, $w_o$ = central peak frequency of the time-resolved spectrum. The observed $w_o$ *generally decreases with time* (Fig.2). *Such narrow-band spectra, if confirmed to be common, are difficult to reconcile with relativistic radiation mechanisms* such as synchrotron or bremsstrahlung, which are intrinsically broadband. Currently there is no general accepted model of FRB emission from magnetars [2] [5] [6], even though some authors have proposed synchrotron maser as a possibility [5] [7]. However, maser-like coherent emission must be narrowly beamed $\Delta\theta \sim 1/\Gamma$ ($\Gamma$=electron beam Lorentz factor), which can be eventually tested with FRB population statistics.

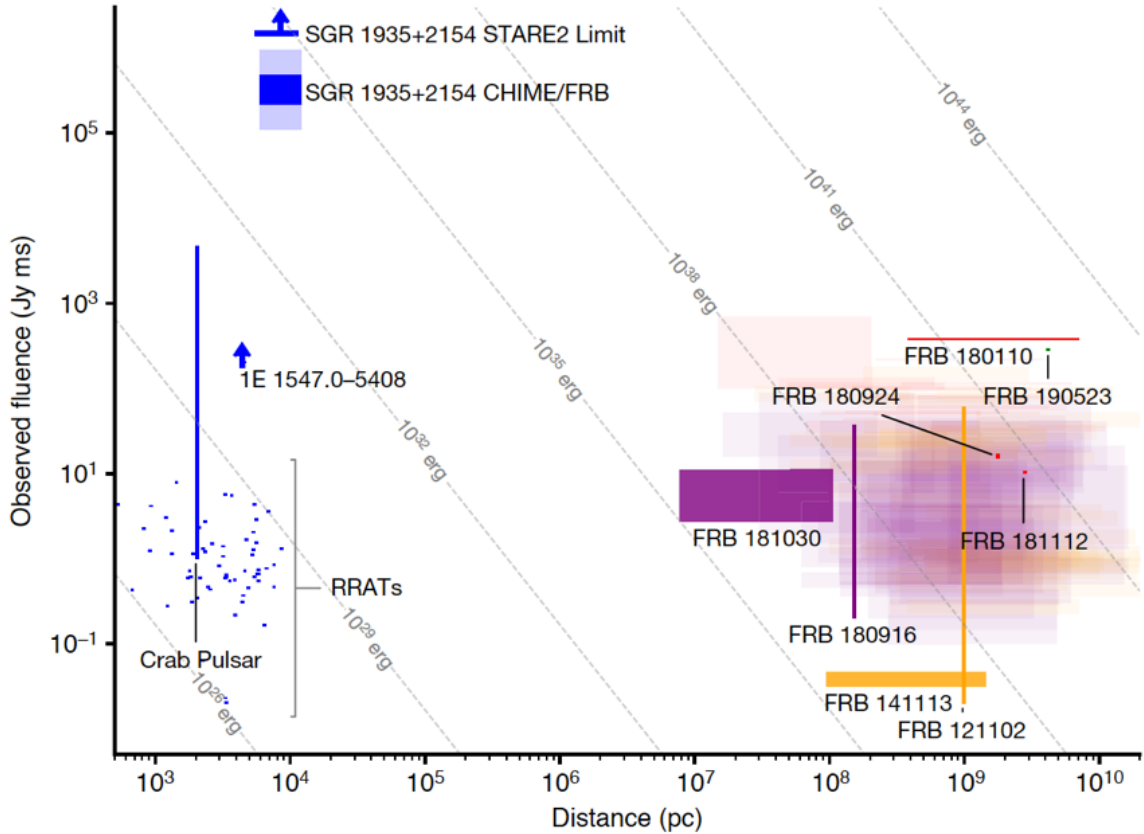

**Fig.1** FRB from SGR1935+2154 flares compared to typical extragalactic FRBs in fluence and isotropic-equivalent energy output $E_{iso}$. Its $E_{iso}$ is within a factor of a few hundred of the weakest burst of repeater FRB 121102, which is also likely from a magnetar. (from [4]).

Among all other known astronomical radio burst phenomena, narrow-band FRBs are most *similar to Type III solar radio bursts (SRB3s), which are also coherent narrow-band emissions with centroid frequencies mostly decreasing with time.* Often two harmonics are observed in SRB3s. The conventional interpretation of SRB3 is that they are plasma line radiation at $w_e$ and $2w_e$ ($w_e$ = electron plasma frequency) emitted by coherent kinetic electron plasma waves (EPW, = Langmuir waves, [8] [9]) via mode conversion by wave-wave scattering, as the disturbance creating the EPWs propagates towards lower densities in the solar corona. One generally accepted driver for these intense EPWs is energetic nonthermal electron beams (e-beam) accelerated by magnetic reconnection. The conversion of propagating e-beam energy into coherent EPWs involve density gradients and plasma instability excited by collective interactions between thermal electrons in the ambient corona and nonthermal beamed electrons from the reconnection sites. Finally, the emission of $w_e$ plasma line radiation arises from scattering of EPW with ion density fluctuations (e.g. ion acoustic waves = IAW), while emission of $2w_e$ line radiation involves EPW-EPW 3-wave scattering [10]. All of these collective plasma processes can only be studied from first principles using kinetic simulations such as particle-in-cell (PIC) simulations [11].

In this paper we will first outline the motivation for pursuing an $w_e$ or $2w_e$ plasma radiation [8] model of narrow-band FRBs from magnetars, and discuss the emission region parameter space constraints of this model from observational data.. Finally we will discuss potential confrontation of these model predictions with multi-wavelength observational data, including x-gamma-ray data
2

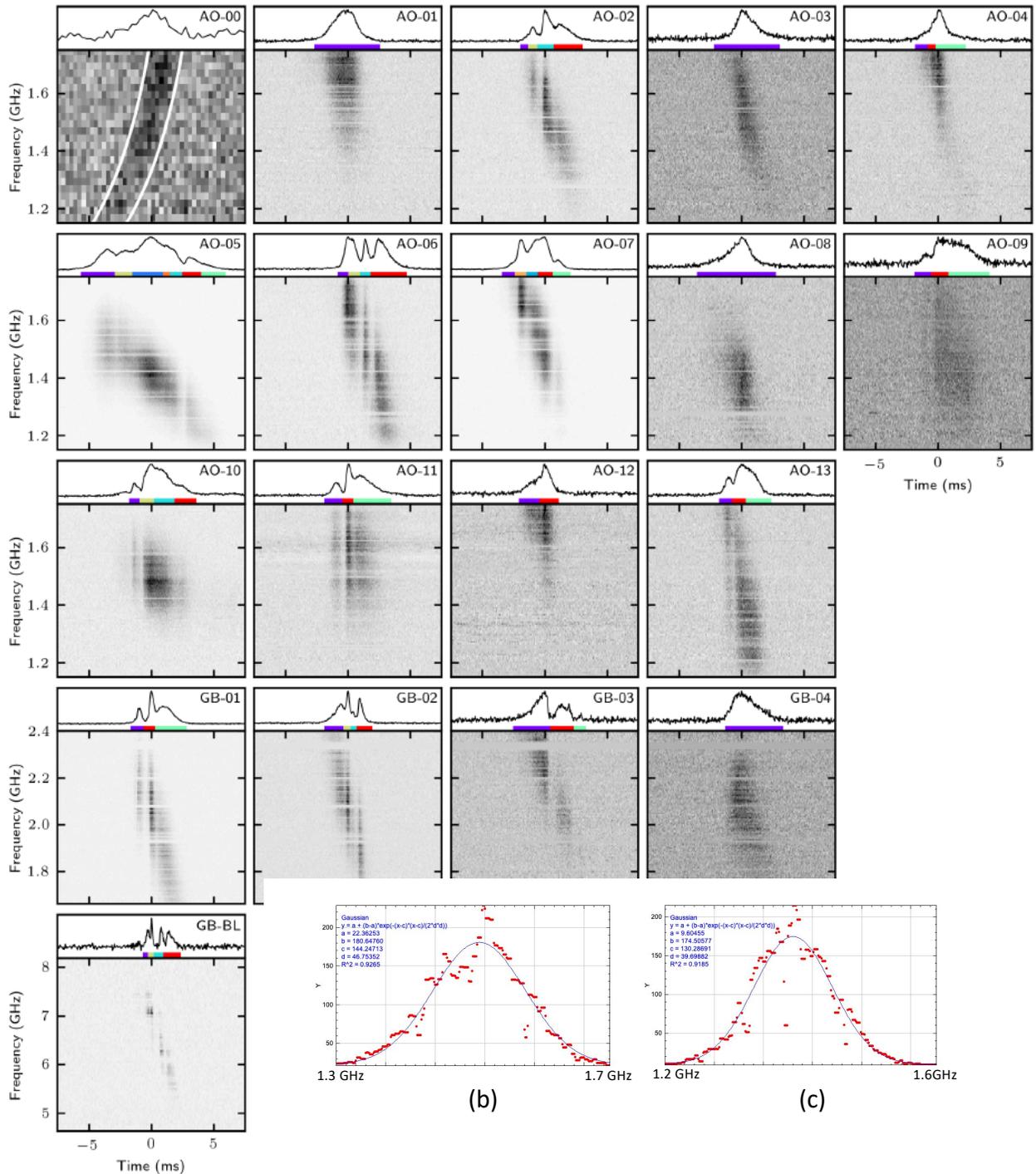

**Fig.2** (a): Dynamic spectra of repeating FRB 121102 bursts (adapted from [12]) showing that the time-resolved spectra are mostly narrow-band with central frequencies decreasing with time. (b)(c): we fit the time-resolved spectra for AO5 and AO10 at peak intensity to highlight the narrow Gaussian line widths (71MHz for AO5 and 87MHz for AO10), which are consistent with narrow line emissions since $\Delta w/w_o < 0.06$.

of magnetars. We emphasize that even a normal star like the sun emits a large variety of radio bursts, and SRB3 is only one of the many types of solar radio bursts [10]. Hence even if most FRBs are associated with magnetar flares, there is likely a large variety of different radio emission mechanisms. *The specific plasma line emission model proposed here is only meant to address those narrow-band FRBs with down-drifting centroid frequency.* Despite this limitation, *our model is attractive because it is highly predictive and readily testable, and the emission region parameters of this model can be meaningfully constrained by observational data,* That is not the case for most other semi-qualitative empirical models (see [2] [5] for reviews). If confirmed, this specific model will provide useful insight into the structure and properties of magnetar magnetosphere, and the dynamics and origins of magnetar flares. Even if this model is disproved, it will still provide useful constraints on the emission mechanisms and origins of FRBs.

## 2. Observational constraints and dimensional estimates of emisssion parameters

Over the past 15 years, a large volume of literature has been devoted to the speculation of the origins and physical mechanisms of FRBs [2] [5]. Unfortunately, even for the recent FRBs which are definitely associated with magnetar flares [4], there is as yet no generally accepted emission mechanism or model. As we mentioned above, we will focus on a specific, highly predictive emission scenario: plasma line radiation at $w_e$ or $2w_e$ emitted by coherent EPWs energized by magnetic reconnection. Because the plasma density is directly related to the observed FRB centroid frequency in this model, the *emission region parameters are tightly constrained by observations*. This paper addresses the basic concepts and order-of-magnitude estimates, while follow-on particle-in-cell (PIC) simulations will provide quantitative results of the kinetic physics [13]. The primary observables of narrow-band FRB's are **(a) duration $\Delta t \sim$ ms, (b) spectrum central frequency $w_o \sim$ GHz (cf. Fig.2), (c) time-integrated fluence of radio and high-energy x-gamma-ray emissions (upper limits in most cases).** When (c) is combined with the distance limits from dispersion measures, we can estimate the isotropic-equivalent radio energy output $E_{iso} \leq 10^{42}$ ergs (Fig.1). This is easily achievable with the known energy range of magnetar outbursts ($\leq 10^{47}$ ergs, [20]). Given the small number of FRBs detected and the coherent nature of the radio emission, it is reasonable to assume that the *radio emission is not isotropic*. Hence the true radio energy output must be reduced by the unknown emission solid angle $\Omega/4\pi$. For convenience of discussion below we define the true radio energy output by $E_R = E_{iso}(\Omega/4\pi)$. If we assume the emission region size $R \leq c*\Delta t$ and identify the electron plasma frequency $w_e$ or $2w_e$ (see below) with the observed $w_o$, conditions (a)(b)(c) above provide meaningful constraints on the emission region parameters (E, n, and $\gamma$), *where E is the average EPW oscillating electric field amplitude, n the average electron density and $\gamma$ the average Lorentz factor of the oscillating electrons in the EPW,* as discussed below. Throughout this paper "electron" refers to both electrons and e+e- pairs which may dominate magnetar magnetospheres.

In order to use conditions (a)(b)(c) to constrain (E, n, $\gamma$), we will adopt three reasonable assumptions for the baseline emission scenario: **(1) the total EPW electric field energy must exceed the emitted radio energy: $(E^2/8\pi)*R^3 \geq E_R$; (2) electron density n and Lorentz factor $\gamma$ are related by $w_o \sim w_e = (4\pi ne^2/\gamma m)$ or $2w_e$, where m is the electron rest mass** (see discussion below); **(3) the Lorentz factor $\gamma$ of the EPW electrons is constrained by equipartition between the electric field energy density and electron kinetic energy density: $E^2/8\pi \sim \gamma nmc^2$.** Applying (1)(2)(3) to conditions (a)(b)(c) we obtain the following dimensional constraints for (E, n and $\gamma$) :

$$\gamma < f*10^4, \quad n < f*10^{14}/cc, \quad E > f*10^6 \text{esu} \tag{1}$$

where $f = k^2(E_R/10^{34}\text{ergs})^{1/2}(\Delta t/\text{ms})^{-3/2}$. Here *we have scaled $E_R$ with the observed $E_{iso}$ of SGR1935+2154* [4]. Note that $E_R$ includes the unknown beam solid angle $\Omega/4\pi$, and $k = 1$ or $2$ depends on whether the FRB centroid frequency is the we or 2we plasma line, since most current GHz radio telescopes cannot observe both plasma lines simultaneously due to their narrow bandwidth (see Sec.6 below). However, condition (c) can in principle be replaced with the observed (or upper limit of undetected) x-gamma radiation output, which is more likely to be quasi-isotropic so that the unknown solid angle can be removed. For the SGR1935+2154 event, the observed x-gamma energy is $\sim 10^{40}$ ergs. This x-gamma energy can be used to replace Eq.(1) with $\Omega \sim 4\pi$ [13].

Besides observables (a)(b)(c), two additional observables are the characteristic (Gaussian) line width $\Delta w$ and the decay rate of $w_o$ (see Fig.2, $\Delta w \sim 71\text{-}87$MHz in these examples, and $w_o$ decrease rate can be extracted from temporal data [12]). $\Delta w$ provides information about the density and energy distribution of the electrons, while the decay rate of $w_o$ provides information on how the magnetospheric electron density decreases with distance away from the magnetar. In the future, if we are lucky to detect both the $w_e$ and $2w_e$ lines simultaneously, then the fluence ratio between the two lines provide additional constraint on the plasma conditions (see Sec.6). Another consideration for the FRBs from SGR1935+2154 is the observed radio-to-gamma-ray ratio of 4 x $10^{-6}$ [12], which is higher than typical FRBs. If we make the reasonable assumption that the x-gamma-ray emission region is near or at the stellar surface, while the radio source is far away in the outer magnetosphere, such ratios provide meaningful *constraints on the conversion efficiency of magnetic reconnection energy to the two kinds of radiation*. Time delay between the radio and high-energy emission will also provide information on the distance from the reconnection site to the two emission regions. A key part of the plasma kinetic physics is to **quantify the conversion efficiency of the input energy from magnetic reconnection (e.g. Alfven waves, electron beams) to $w_e$ and $2w_e$ plasma line emissions from first principles using PIC simulations [13]**.

### 3. Kinetic physics of $w_e$ and $2w_e$ plasma line radiation

It has been demonstrated that enhanced we and 2we plasma line radiation will be emitted by electron plasma waves (EPWs) when there is steep density gradient or the presence of nonthermal hot electrons in a cold plasma [10] [14] [15] [16]. *The generally accepteed paradigm is that the $w_e$ line radiation is emitted by EPW scattering from ion density fluctuations and the $2w_e$ radiation is emitted by EPW-EPW (three-wave) scattering.* In solar cases, the above conditions arise from e-beams ascending the solar corona after they are produced by magnetic reconnection at lower altitude [10] [17]. Here we propose that such conditions are also likely to prevail in magnetar flare regions in the magnetosphere. Magnetic reconnection in the magnetar magnetosphere will dump a large amount of nonthermal energy in the form of accelerated particles and Alfven waves which propagating along field lines *both towards the stellar surface and away from the magnetar* (see Sec.6 below). In the former case the energy flux heads towards higher magnetic fields and higher density. There it will efficiently convert into x-gamma radiation. The oppositely directed energy flux heads toward lower fields and lower density in the outer magnetosphere. There some of it will dissipate via collisionless processes into coherent electron plasma oscillations which can then radiate $w_e$ and $2w_e$ plasma line radiation. We propose that this is at least one plausible emission mechanism of the FRBs from magnetars. What is unknown is the *efficiency of this mechanism in the magnetar setting, which may be radically different from the solar corona.*

In the solar SRB3 case, the EPWs couple to $w_e$ radiation through a steep density gradient of scale height L ($> 2\pi d_e$, $d_e=c/w_e$), from [8] the $w_e$ line emissivity is estimated as $2cd_e |E \sin\theta/L|^2$ where $\theta$ is the angle between **grad** n and **E**. Assuming $<\sin^2\theta> = 1/2$, the limit on E from Eq.(1) then gives a useful estimate of the FRB radio flux as a function of L. Thus, if we believe that the x-gamma-ray flare of a magnetar originates from dense plasmas at or near the stellar surface energized by reconnection energy flux propagating downward, then the flare should also emit significant level of EPW flux propagating upward, and some of it will convert to $w_e$ and $2w_e$ plasma line radiation in the outer magnetosphere. The minimum luminosity in these lines would be $\sim 10^{-6}$ to $10^{-8}$ of the x-gamma luminosity. The plasma line radio luminosity could go much higher if the radio and x-gamma emitting site is coincident with the radio site, and the x-gamma-emitting electrons are heated by the EPW themselves, as in laser plasma interactions [18] in the laboratory.

The far-reaching implications of the identification of narrow-band FRB emission with plasma line radiations are manifold. Besides providing a valuable diagnostic of the FRB emission region, the plasma physics of efficient EPW excitation and mode conversion is extremely interesting in its own right. Comprehensive understanding of these kinetic phenomena represents major advance in fundamental plasma physics, and will likely lead to many other applications besides high-energy astrophysics, such as the diagnostic of laboratory plasmas, solar and space physics.

The next question is how the EPWs are efficiently generated by energy fluxes coming out of the magnetic reconnection region. The coupling of electron beams to EPWs via the strong electric fields is well understood from both experiments and PIC simulations, so we will not dwell on it here. But the coupling of relativistic Alfven waves to EPWs needs a little elaboration. Relativistic Alfven waves (RAW) with $B/(4\pi\rho)^{1/2} > c$ behave very much like vacuum EM waves along a guide field except for the small plasma loading. When a RAW is incident on a steep density gradient, it will generate nonlinear EPWs near the "critical surface" where the electron plasma frequency is equal to the RAW wave frequency. A dominant mechanism for EPW generation near the critical surface is resonant absorption [16] [18].

The basic concept of resonant absorption to produce EPWs can be understood as follows [18]. When an intense flux of transverse wave (e.g. relativistic Alfven wave) of characteristic frequency $w_A$ is incident on a density boundary, penetrating beyond the critical density, nonlinear EPWs (Langmuir waves) will be excited as long as the incident RAW electric field $\delta E \sim \delta B$ has a component parallel to the density gradient **grad** n. Nonlinear EPWs propagating down the density gradient will break (as in the case of shallow water waves), dumping their energy into superthermal or hot electrons. But those EPWs propagating into higher density regions will not break. Some EPW will dissipate via Landau damping [8] [11]. But some EPW will scatter with ambient fluctuations to emit $w_e$ and $2w_e$ line radiation. Hence it is natural to expect that collisionless absorption of RAW may lead to efficient excitation of EPWs and subsequent emission of $w_e$ and $2w_e$ plasma line radiation. *The main unknown is the conversion efficiency of RAW into EPWs.* It is important to note that when a highly nonlinear RAW is propagating into lower ambient density regions, the Poynting flux pressure can in principle "snowplow" and compress the upstream plasma via **jxB** force into higher density plasma so that an initially "underdense" plasma eventually becomes "overdense", and capable of mode conversion of RAW into EPWs.

## 4. Generalization of plasma line radiation mechanisms to magnetar regimes

As we discussed above, our model of FRB in the context of magnetar flares is based on the concept of mode conversion from longitudinal EPWs to transverse EM waves at $w_e$ and $2w_e$ frequencies. In the case of solar SRB3's, the primary energy source is believed to be magnetic reconnection. Reconnection ejects high energy nonthermal electron beams which propagate along field lines away from the sun. A combination of density gradient effects and the interaction of ambient thermal plasma with the nonthermal electron beams leads to the excitation of coherent EPWs. Finally, mode conversion of coherent EPWs via wave scattering leads to the copious emission of transverse EM waves centered at $w_e$ and $2w_e$ frequencies.

To extend the above solar model to the setting of magnetar flares, we propose to start with a conceptual framework based on SRB3s, but *adding new ingredients unique to the magnetar flare regimes*. The first new ingredient will be strong B fields so that *Alfven waves will likely carry energy flux comparable to or exceeding electron beams from the reconnection site. Such Alfven waves are expected to be highly relativistic $(B/(4\pi\rho)^{1/2} > c$ and $w_L > w_e)$*. The second new ingredient will be that the *electron momentum in the EPW oscillations will be highly relativistic with $\gamma >> 1$* (cf. Section 1.1). The third new ingredient will be the likely *dominance of e+e- pairs over ions*, such that $n_+$ *(positron density)* $>$ or $>> n_i$ *(ion density)*. In this case we expect the *$2w_e$ line to dominate the $w_e$ line since IAW excitations will be small compared to EPW excitations*. Conventional picture of a quiescent magnetar magnetosphere is that it should have little ion content. However, during strong flares ions may be driven from the surface to the magnetosphere, so that the ion content in the outer magnetosphere may not be negligible. To cover all possible scenarios, we need to carry out PIC simulations with a range of $n_+/n_i$ ratios [13]. *When $n_+/n_i$ is low, we expect IAW or ion fluctuations to be abundant, so that the $w_e$ line should dominate over the $2w_e$ line as in most SRB3s. But when $n_+/n_i$ is high, IAW should be negligible compared to EPW, and the $w_e$ line would be suppressed relative to the $2w_e$ line.* Another important difference caused by the very strong magnetic fields in the magnetar magnetosphere is that we expect the *longitudinal EPWs to mainly propagate along field lines*. Because the ion cyclotron and ion plasma frequencies are both $<< w_e$, the $w_e$ radiation is, to the first order, unaffected by the B-field. But we expect the $2w_e$ radiation to be emitted only where the B field is curved since strictly co-linear EPW-EPW scattering cannot directly produce transverse EM waves. Alternatively we may resort to higher order processes such as scattering of transverse $w_e$ radiation with longitudinal EPWs and so on. Of course, the EPWs can also scatter from the other collective modes not propagating parallel to B (for example, lower hybrid resonance [8]). But all these waves have frequencies $\sim w_L = eB/mc$. Hence the resultant electromagnetic radiation will have characteristic frequency near the electron cyclotron frequency, making them indistinguishable from ordinary cyclotron emission. Hence the *presence or absence of both plasma lines and their ratio will be a useful diagnostics of the emission region* [16] (see below). However we should point out that the deficit of IAW in e+e- pair plasmas does not automatically rule out the emission of the first harmonic $w_e$ line. Scattering of EPW with transverse waves can also convert EPW into $w_e$ radiation and bootstrap the net efficiency of emergent $w_e$ radiation.

**To our knowledge, the generation and mode conversion of nonlinear relativistic EPWs in strongly magnetized pair-dominated plasmas have not been systematically studied using PIC simulations.** Hence it is important to systematically study the basic plasma physics of EPW excitation and mode conversion in these exotic regimes of magnetars, using PIC simulations. Only after we have achieved a thorough understanding of EPW generation and mode conversions in such regimes can we apply these kinetic results to construct more empirical global astrophysical

models and make observational predictions. In Sec.6 we discuss the diagnostic applications of the line ratio.

## 5. Platforms and setups for PIC simulations

In this paper we first outline the platforms and setups for PIC simulations of EPW excitation and radiation mechanisms. Detailed PIC simulation results will be presented in a follow-on paper [13]. While some PIC simulations of plasma line radiation by EPW have been studied before in the context of SRB3s, we will focus on new plasma regimes relevant to magnetar magnetospheres which include relativistic electrons and e+e- pairs and strong ambient magnetic fields. Fig.3 illustrates a typical 3D PIC box setup. we will first perform 2D runs in the x-y and y-z planes before pursuing 3D runs. we will set up the initial EPWs at t=0 with 2 different numerical methods, by

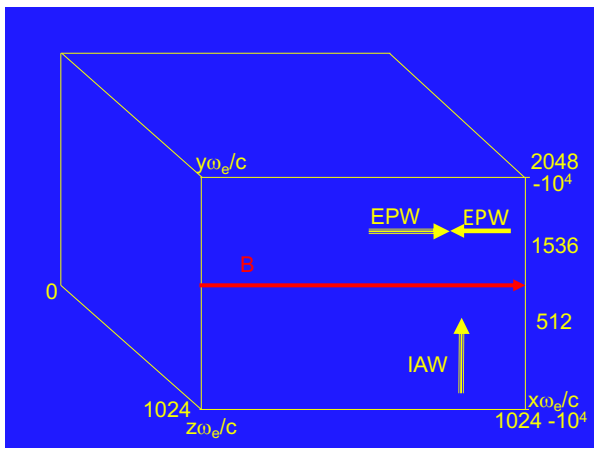

**Fig.3** Generic 3D PIC box of proposed PIC simulations. Both EPWs and guide field **B** will be set up along the x-axis. IAWs or secondary EPWs will be injected ain the x-y plane at different angles to study wave-wave scattering to produce we and 2we radiation. we start with 2D runs in the x-y, y-z, and x-z planes before doing 3D runs to cross-check and validate the 2D results.

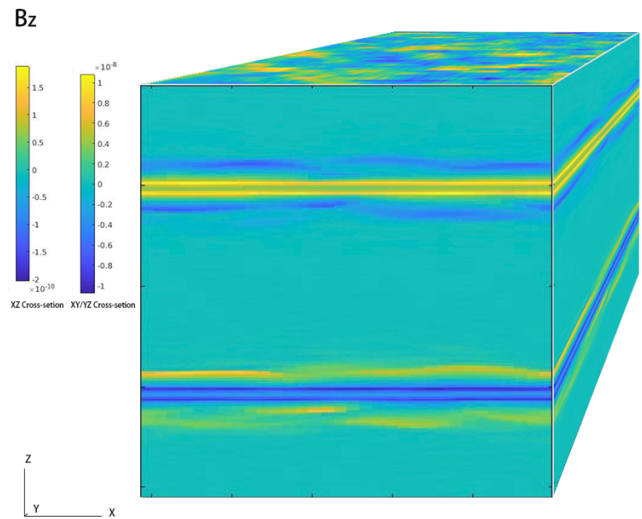

**Fig.4** Sample 3D EPOCH PIC output of a relativistic shear flow problem with 1024x256x2048 cells using periodic boundaries. The bulk Lorentz factors of the two opposing flows = +/- 15. The transition layers have scale heights of L=50 cells. we plot the $B_z$ contours here to show the double slab structure at the shear boundary. This result validates 2D results done in the x-y and y-z planes (from [19], here y and z are interchanged from those of Fig.3).

(a) applying a transient electric field, or (b) loading the particles with initial nonzero charge separation. We will then inject low-frequency ion acoustic waves (IAWs) or counter-propagating EPWs into the PIC box (Fig.3). To study the efficiency of wave-wave interactions leading to $w_e$ and $2w_e$ line radiations, and characterize the properties of the transverse EM waves, we will systematically vary the following parameters: (1) injection angle of IAW/EPW with respect to ambient EPW, (2) amplitudes of both the IAW and EPW, (3) temperatures of the electrons, pairs and ions, (4) $n_+/n_i$ ratio, (5) amplitude of the guide field B. Fig.3 illustrates the generic setup. Fig.4 showcases a sample 3D EPOCH output of a relativistic shear flow run used to validate much larger 2D runs in the x-y and y-z planes [19]. The 3D box size was 1024x2048x256 cells with ~ $10^{10}$ particles.

In the absence of an ambient guide field to break the symmetry, 2D simulations can fully capture the essential physics of EPW-IAW and EPW-EPW interactions and plasma line radiation. Once an ambient B field is added to the problem, 2D simulations will be limited to represent only very special cases. 3D simulations will be needed to investigate the full range of interactions. Nonetheless, we can still start with some 2D runs before expanding to 3D runs. For example, if either the EPW or the IAW is parallel to the B field, then 2D suffices. If both the EPW and the IAW are orthogonal to the B field, then 2D also suffices. But when both the EPW and IAW are oblique to the B field, then 3D must be used. we emphasize that the plasma line width in our case depends on both density gradient and electron temperature, which itself can be highly relativistic.

Next we will simulate $2w_e$ plasma line radiation from EPW-EPW scattering. In the absence of a background B field, such simulations are identical to the EPW-IAW scattering runs above, substituting IAW with a second EPW. 2D runs should be able to capture the essential physics in such cases. However, when a strong ambient B field is added to the problem, EPW is mainly propagating parallel to the B field, while transverse EPW's will be suppressed [16]. However, head-on collisions between two strictly co-linear EPWs cannot produce transverse EM waves [16]. A more promising candidate is the scattering of EPW with transverse EM modes, which can propagate in any direction relative to the B field. we can carry out such simulations in both 2D and 3D. As we noted above, in a pair-dominated plasma, IAW should be negligible compared to EPW. Hence we expect the $2w_e$ line to dominate the $w_e$ line. This prediction is testable with future observations (see Sec.6).

For SRB3 emissions from the sun, most models invoke hot electron beams (e-beams) emanating from magnetic reconnection sites. There are many spatially resolved multi-wavelength solar observations to support this picture [10]. However, e-beams are charged and the super-high currents required by the FRB energetics make the e-beam scenario challenging in the case of magnetars. At the same time, magnetars have strong magnetic fields even in the ambient magnetopshere. Any reconnection scenario will likely convert a large fraction of energy released into relativistic Alfven waves in addition to accelerated particles. Hence we need to investigate several scenarios of EPW excitation in magnetars, different from those of SRB3's as discussed below.

First we need to extend the e-beam excitation scenario of SRB3 to the relativistic regime with e-beam bulk Lorentz factors $\Gamma$ and internal $\gamma$ both >> 1. To study the coupling of relativistic e-beams to EPWs in a cold plasma, we start with the benchmark case with no ambient magnetic field, using mixed-boundary PIC simulation boxes: the incident e-beam direction (x-axis of Fig.3) will have open boundary while the y-axis will have periodic boundary. We need to study the properties of EPWs created by the e-beam by varying the e-beam Lorentz factors, current density and ambient plasma temperatures. Next we need to add e+e- pairs to both the e-beam and the ambient plasma and vary the pair ratio ($n_+/n_i$). We need to study EPW generation as we vary both the magnitude and direction (i.e. pitch angle) of ambient **B** relative to the e-beam direction. The physics of EPW excitation can be adequately studied in 2D as long as both **B** and plasma density are uniform in the z-dimension. These PIC simulations can be calibrated and validated using SRB3 results.

Next we need to study the excitation of EPWs by nonlinear ($\delta B \geq B$) relativistic ($v_A \geq c$) Alfven waves (RAW). It is well known in laser plasma interactions [18] that large amplitude EM waves (e.g. linearly polarized laser) can strongly couple to EPWs near the critical density surface [16] via collective processes such as resonant absorption [18] [16]. Relativistic Alfven waves are similar to EM waves propagating along a guide field. Hence we expect RAWs will also efficiently transfer their energy to EPWs at the critical density. The energy transfer efficiency from lasers to EPWs can be as high as 50% [18]. Hence in magnetars RAWs can be a much more effective agent of energy transfer than electron beams from the magnetic reconnection site to the radio emission site.

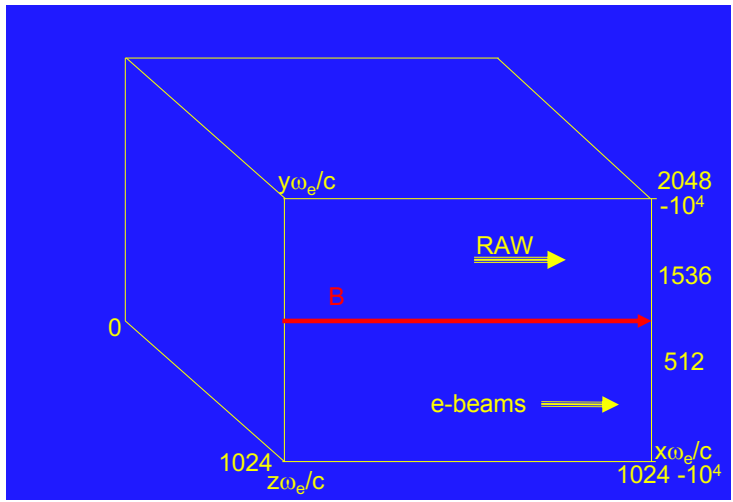

**Fig.5** Generic PIC box to illustrate the simulations of RAW and e-beam excitation of EPWs in an ambient plasma. In this case open boundaries will be used in the direction of flux injection. In general, the guide field B will not be parallel to the flux injection direction. If both RAW and e-beams are injected and they are not in the same direction as the guide field, 3D simulation is required.

## 6. Astrophysical models and discussions

Using PIC simulation results on the efficiency of different energy transfer mechanisms and plasma line radiation processes, we will develop a concrete plasma line radiation scenario of FRB's from magnetars. The plan is to arrive at a series of **concrete physical models with "clean" observable predictions that can be systematically tested with observational data**. The goals are twofold: (a) to critically test the basic viability of the plasma line radiation picture; (b) if the answer to (a) is positive, then to use the observational data to discriminate between specific empirical models of $w_e$ radiation (e.g. e-beam vs. RAW energization of EPW, strongly magnetized vs. weakly magnetized emission region, pair to ion ratio, etc). As we discussed in Sec.1, currently there is no firm observational evidence of two harmonic emissions among most FRBs. If confirmed this may support the existence of strong fields and / or ion deficiency, in the emission region (cf. Sec.1). The PIC simulation results will be useful for such studies. Conversely, if future broad-band radio observations confirm the presence of both lines, then the line ratio will be useful to further constrain the emission region parameters and emission mechanisms (see Sec.2.3 below).

Besides the observational constraints discussed in Sec.1, another important data set will be polarization. For example the repeating FRB 121102 shows ~100% linear polarization with large Faraday rotation measure, and ~ 0% circular polarization [12]. If such data are confirmed for other FRBs from known magnetars, this can be used to further constrain the emission parameters based on our PIC simulation results [13], which should predict the polarization.

Fig.6 shows a schematic cartoon of the FRB emission scenario. Magnetic reconnection in the inner magnetosphere can be triggered by newly emergent flux from the magnetar surface stressing existing magnetic fields similar to solar flares. Reconnection releases large amount of energy which leave the reconnection site in two generally opposite energy fluxes: one propagating downward towards the magnetar surface while the other one propagating outward into the outer magnetosphere. The downward energy flux gets mostly converted into heat and x-gamma radiation [20] via bremsstrahlung, cyclo-synchrotron and inverse Compton scattering [21], with the optically thick region emitting blackbody-like radiation. The upward flux propagates into lower plasma density and decreasing magnetic fields. Hence it should be less radiative in high-energy radiation [20], but more likely to emit radio emission due to weaker fields and lower density. Assuming that the initial reconnection input into the two opposite energy fluxes are comparable, which seems likely, then the observed ratio (or upper limits) of radio to high-energy radiation [20] largely depends on the conversion efficiency of primary energy flux to radio emission. The super-high brightness and coherence of the radio emission leave very few viable options for the radio emission mechanisms, such as $w_e$ plasma radiation from coherent EPWs, or relativistic coherent synchrotron-maser action similar to the free-electron laser. We will examine in-depth the observable differences between these two mechanisms. Among all viable scenarios of magnetar FRBs, we believe that the *plasma line radiation model is most predictive and therefore most testable with observational data.*

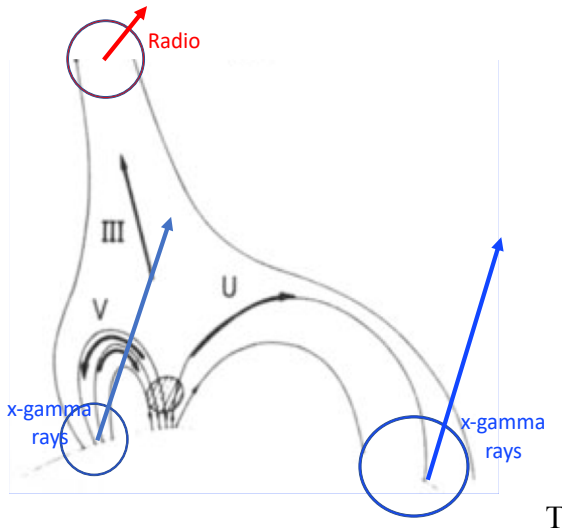

**Fig.6** Schematic sketch of magnetic reconnection near a Y-branch magnetic field configuration so that accelerated electron beams and Alfven wave fluxes emitted by the reconnection can travel both down to the stellar surface via closed field lines, and up to the outer magnetosphere along open field lines (from [17]). The time delay and relative energy outputs between the near-surface emission and the outer magnetospheric emission provide useful constraints on the distance of the reconnection site to the two emission regions and radiation conversion efficiencies [13].

As far as testing the plasma line radiations scenario, one of the most important test is the presence or absence of the two harmonics emission: presence of both harmonics will be a strong confirmation. Due to the narrow bandwidth of current radio telescopes, it is inherently difficult to observe two harmonics at ~ GHz frequencies even if they are present, especially if they drift in time. However, in rare fortuitous cases this may occur. Figure 7 shows the tantalizing spectra of FRB 121102A with a *hint* of both $w_e$ and $2w_e$ line presence, since *the peak frequency of the first burst is close to twice the peak frequency of the second burst.* But the telescope bandwidth is not able to cover both lines sufficiently to firmly prove this hypothesis. One future strategy may be to use two telescopes whose frequency bands differ by a factor ~ two for simultaneous observations. If the two peaks in Fig.7 indeed correspond to two harmonics, the higher harmonic flux and fluence are ~ twice the flux and fluence of the lower harmonic, suggesting some suppression of IAW relative to EPW in the emission region, consistent with e+e- pair domination or ion deficiency.

Another important observational diagnotic is to measure the polarization, both linear and circular, in all cases.

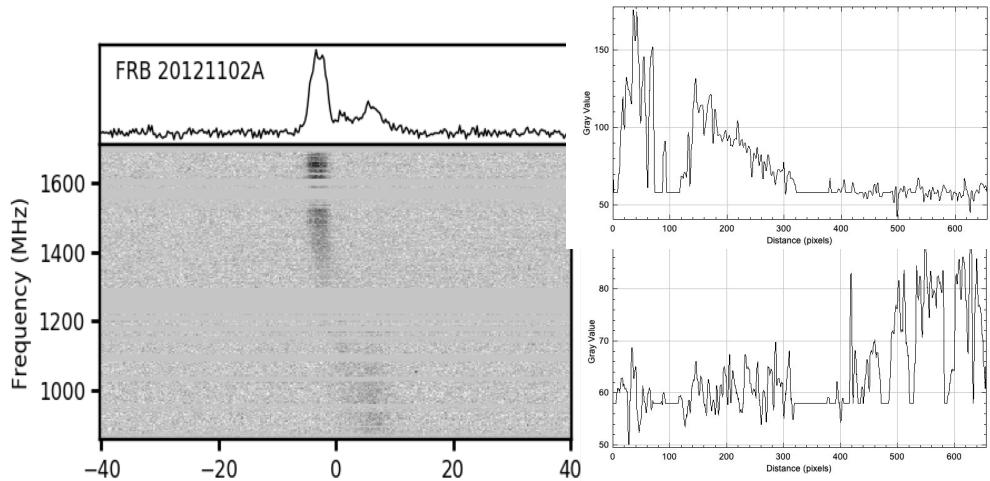

**Fig.7** (left) Time resolved spectra of consecutive bursts from FRB121102A show hints of two frequency peaks in the ratio of 2:1 [from 22]. (right) Lineouts of the FRB intensities at peak flux of the first and third peaks. The x-scale of the line-outs taken at peak intensity of each peak coincides with the frequency range of the left picture. The peak intensity and fluence of the high frequency peak are ~ twice of the low frequency peak.